\input epsf.sty

\documentstyle[12pt,aasms]{article}

\slugcomment{Submitted Ap. J. Letters }

\begin{document}

\title{The Anisotropy in the  Cosmic Microwave Background\\
       At Degree Angular Scales.}

\author{C. B. Netterfield, N. Jarosik, L. Page,
        D. Wilkinson, \& E. Wollack \altaffilmark{1}}
\affil{Princeton University, Department of Physics, Jadwin Hall,
       P.O. Box 708, Princeton, NJ 08544}


\altaffiltext{1}{NRAO, 2015 Ivy Rd., Charlottesville, VA, 22903}


\begin{abstract}

	We detect anisotropy in the cosmic microwave background
(CMB) at degree angular scales and confirm a previous detection
reported by Wollack et al. (1993). The root-mean-squared amplitude
of the fluctuations is $44^{+13}_{-7} \mu$K. This may be expressed as the
square root of the angular power spectrum in a band of multipoles
between  $l_{eff} =69^{+29}_{-22}$. We find
$\delta T_l = \sqrt{l(2l+1)<|a_l^m|^2>/4\pi} = 42^{+12}_{-7} \mu$K.
The measured spectral index of the fluctuations
is consistent with zero, the value expected for the CMB. The spectral index
corresponding to  Galactic free-free emission, the most likely foreground
contaminant, is rejected at approximately $3\sigma$.

	The analysis is based on three independent data sets. The
first, taken in 1993, spans the 26 - 36 GHz frequency range with
three frequency bands; the second was taken with the same
radiometer as the first but during an independent observing
campaign in 1994; and the third, also take in 1994, spans the
36-46 GHz range in three bands. For each telescope
position and radiometer channel, the drifts in the
instrument offset are $\le 4~\mu$K/day over a period of
one month.  The dependence of the inferred anisotropy on the
calibration and data editing is addressed.

\end{abstract}


%
%

\section{Introduction}

	In Wollack et al. (1993), we reported on a measurement
of the anisotropy in the CMB made from Saskatoon, SK (SK93). The
results were most parsimoniously described as a detection of CMB
anisotropy. However, an anisotropy in Galactic free-free emission
could have produced similar data with $\approx 10$\% probability.
We have since returned to Saskatoon and observed the
same fields  with an  an extended frequency baseline that now
spans 26 to 46 GHz. The new data, SK94, confirm the 1993  results
and strongly constrain the spectral index of the fluctuations.
Reviews of the state of CMB anisotropy measurements are given in
Bond (1994),  Readhead \& Lawrence (1992), and White et al.
(1994).

\section{The Instrument and Observing Strategy}

	Observations were made with two independent radiometers,
one in ${\rm K_a}$ band and one in ${\rm Q}$ band, that
sequentially shared the same telescope. The details of the
instrument may be found in Wollack (1994a, 1994b), Wollack et al. (1993,
hereafter W1) and Page et al. (1994). Each radiometer
detects the total power in three frequency bands and two
polarizations\footnote{A cracked waveguide joint vitiated the
data from the vertical polarization in the ${\rm Q}$ radiometer;
it is not used.}. The radiation is amplified by HEMTs
(Pospieszalski et al. 1988; Pospieszalski 1992 )  cooled to 15K.
The passbands for ${\rm K_a}$ band are 26-29, 29-32, 32-36 GHz,
and for ${\rm Q}$ band are 36-39.5, 39.5-43, 43-46 GHz. The
six channels in each radiometer view the sky in a single
approximately frequency-independent beam formed by a cold
corrugated horn feeding an ambient temperature off-axis parabola.
The FWHM beam in ${\rm K_a}$ band is $1.42\pm 0.02^{\circ}$ and in
${\rm Q}$ band is $1.04\pm 0.02^{\circ}$. After the parabola,
the beam reflects off of a 90 cm$\times$ 150 cm chopping flat
that scans the beam back and forth about a vertical axis, following an
approximately triangular
waveform at 3.90625 Hz. A large stationary ground screen shields the
radiometer from terrestrial and solar emission. Other than the
new radiometer and the modulation scheme  described below, the
principle improvements to the instrument described in W1 are 1)
an enlarged ground screen; in our data, contaminating signals
from the ground and sun are less than $1~\mu$K.
2) thermal stabilization of all radiometer components
to better than 0.1 K; and 3) increased vibrational isolation of
the radiometers.

	The chopping flat sweeps the beam in azimuth while the
radiometer channels are synchronously sampled at 250 Hz. The
the angle between the sweep endpoints subtends
$7.01^{\circ}$ and $7.36^{\circ}$  for the ${\rm K_a}$ and  ${\rm
Q}$ systems respectively; thus, the $3\,$dB antenna beam contours of the two
systems match at the extrema of their sweeps. The center of the
sweep pattern is wobbled every 20 seconds from $4.42^{\circ}$ west of the north
celestial pole to $4.42^{\circ}$ east.

	The effective beam pattern is synthesized in software by
appropriately weighting each sample during a chopper sweep. This
allows us to probe multiple angular scales, optimize the spatial
frequency coverage to mitigate atmospheric contamination, examine
the quadrature demodulation, and match the ${\rm K_a}$ and ${\rm
Q}$ synthesized profiles even though the FWHMs are different.
The full analysis based on this scheme will be discussed in a
forthcoming paper; here, we use the method to synthesize the SK93
pattern. The beam pattern for one chopper sweep is shown in
Figure 1. The data are normalized so that a $1~\mu$K temperature
change in the positive lobe of a beam averaged over $15^{\circ}$
in right ascension (1 bin) gives a signal of $1~\mu$K. A good
approximation to the window function is
$W_l=[1.5-2P_l(cos[\theta_t])+.5P_l(cos[2\theta_t])]e^{l(l+1)\sigma^2}$
where $P_l$ are the Legendre polynomials, $\theta_t =
2.57^{\circ}$ and $\sigma=0.64^{\circ}$.

	The instrument is calibrated using Cassiopeia-A.  From a
fit of the 8-250 GHz data in Baars (1979) and Metzger (1986), we
find $S_{\nu} = (2070\pm 162)\nu^{(-0.695\pm0.029)}$ Jy,
($\nu$ in GHz, epoch 1994)  corresponding to a  $9.57\pm1.2~$mK
signal with a $1.04^{\circ}$ beam at  38.3 GHz.  The laboratory
calibrations agree with this to within 20\%. Because Cas-A is
observed in the same thermal  and radiometric environment in
which the CMB is observed, we consider it the definitive
calibration. Cas-A is also used to map the beam; the values are
within $0.02^{\circ}$ of those found with a microwave source. The
absolute pointing error, as determined from Cas-A, is
$0.06^{\circ}$.

\section{Data Reduction and Analysis}

	In essence, the data reduction consists of editing as
described below, binning the results on the sky, and subtracting
the mean of the binned data. The results are shown in Figure 2.
Although the constant part of the  demodulated
signal, or offset, depends on the east/west position and on the
frequency and polarization, it is relatively independent of time. A typical
drift is $3\pm3~\mu$K/day in both radiometers.
Variations in the chopping flat  temperature will effect the offset
because of the dependence of the emissivity on the incident
angle. The offset is fitted to the temperature of the chopper
and a correlation is found; however, removing the small
variation has a negligible effect on the final data set.

	There are three components to the data editing procedure.
1) The data from the single-difference demodulation
\footnote{``single-difference'' refers to the demodulation where
the data taken with the chopper pointed west are subtracted from
those where the chopper is pointing east.}, which is sensitive to
horizontal spatial gradients in the sky, are used to monitor the
atmosphere. If the mean deviation of 8s averages of data is
greater than some value, $\eta$, then the 15 minute section
containing those data is cut. In Table 1, we have converted
$\eta$ into the standard deviation of the variation in the
horizontal thermal gradient of the atmosphere, $\zeta$. This
value is relatively independent of the observing strategy. In
addition, for a 15 minute section to be included, both abutting
15 minute sections must also be good. 2) Next, the standard
deviation of the 64 datum from each chopper  cycle is computed.
If one or more points deviates by 3.5$\sigma$, then the entire
chopper cycle is deleted. This cuts an additional 2.6\% of the
good data. 3) Finally, we require that no 20s average of data in
a right-ascension bin deviate by more than $3.5\sigma$ of all the
data in the bin. This cut rejects $\approx 1.5$\% of the data
that passed the atmosphere cut and  the final results are robust
to variations in this cut level.

	The data quality is monitored with a quadrature
demodulation (data taken during the clockwise chopper scan
minus data taken during the counterclockwise scan). This signal
is consistent with zero.

	To quantify the anisotropy signal, we use the Bayesian method
described in W1 to find the root-mean-squared amplitude,
$\Delta_{rms}$ and index, $\beta$, of the fluctuations. In the
full implementation, the new data are  analyzed along with the
SK93 data using a $720\times 720$ correlation  matrix. A maximum
of the likelihood, $L(\Delta_{rms},\beta)$, is found when
$\Delta_{rms}$ and $\beta$ scale the fixed spatial correlations
inherent in the observing pattern so that they best describe the
patterns in the measured data.  The data taken in the east and
west are treated independently and the atmosphere and HEMT
induced correlations are accounted for (Dodelson \& Kosowsky
1994a).  The resultant $\Delta_{rms}$ depends on the observing
strategy. We convert it into an estimate of square root of the
angular power spectrum in a band of multipoles between $l=47$ and
$l=98$ ($l_{eff} = 69^{+29}_{-22}$ ) following Bond (1994) and
Peebles (1994). The results are given in Table 1.

\clearpage
\section {Discussion}

	Table 1 shows that there is a strong detection of
anisotropy with $\beta\approx0$, independent of atmosphere cut
and data sub-set.  If the ${\rm K_a93}$ data are subtracted from
the  ${\rm K_a94}$ data, the result is consistent with zero, as
expected if both see the same signal. In other words, the ${\rm
K_a94}$ results confirm the  ${\rm K_a93}$ results.

	There are two dominant sources  of systematic error in
this measurement; inaccuracy in our knowledge of the flux from
Cas-A (13\%) and an inaccuracy in our measurement of Cas-A
(Q:~6\%, Ka:~3\%). Small errors are introduced by inaccuracies in
the beam FWHM  (2\%), in the measurements of the passbands
($2\%$),  and in the phases of the recorded signals with respect
to the optical axis ($<1\%$). The combination of all these errors
leads to a $\pm 15\%$ uncertainty in our temperature scale.  For
the best estimate of celestial fluctuations we give the result
with the most statistical weight ($44^{+11}_{-6}~\mu$K) with the
$\pm 15\%$ added in quadrature, $\Delta_{rms} =
44^{+13}_{-7}~\mu$K. Converting to   a band-power\footnote{The
slight differences between Peebles (1994) and Bond (1994) may be
ignored for large multipoles, $l$.}, we find $\delta
T_l=\sqrt{l(2l+1)<|a_l^m|^2>/4\pi} = 42^{+12}_{-7} \mu$K at
$l_{eff} = 69^{+29}_{-22}$.

	In Table 1, one notes that the addition of new data did
not improve the statistical error bar by the square root of the
observing time. In part,
this is because the error on $\Delta_{rms}$ contains the sample
variance, which is a function of the area of the sky observed. The
statistical error bar for the ${\rm K_a93}$ data is $\approx 10.5
{}~\mu$K (the expectation value, opposed to the maximum of the
likelihood,  is roughly half way between the 16\% and 84\%
integrals). If the sample variance, which is computed to be $6.5
{}~\mu$K, is subtracted from this in quadrature then one finds an
``intrinsic'' error of $8.2~\mu$K. The full data set should have
an intrinsic error of $8.2 \sqrt{130{\rm h}/428{\rm h}~\mu}K
= 4.5~\mu$K which is
close to the computed value of $5.5~\mu$K. In addition, by
inspection of Figure 2 and Table 1, one sees that the data from the three
radiometers may differ by more than one would expect from
random noise alone, possibly due to the atmosphere. This would
tend broaden $L(\Delta_{rms},\beta)$.

	There are two sources of a systematic error on $\beta$,
uncertainty in the relative calibration between channels
and the uncertainty in the fit to the Cas-A calibration data . These combine
to produce an error
of $\pm 0.08$. In addition, we cannot be certain that atmospheric
fluctuations are
not biasing our answer. While the bottom two entries in Table 1
agree statistically, one might worry that low cut levels result
in more negative indices. To ensure that the quoted value for
$\beta$ represents one's confidence, we quote the
average value of $\beta$ with an error bar that nearly
encompasses the errors from both entries. Including the
systematic uncertainty,  $\beta=-0.1^{+0.8}_{-0.8}$.

	To investigate the possibility that $\Delta_{rms}$ is due
to a Galactic foreground, we examine $L(\Delta_{rms},\beta)$
bearing in mind that there might be an atmospheric bias.
$L(\Delta_{rms},\beta)$ is normalized to 1 at its maximum. For
free-free emission, $max[L(\Delta_{rms},-2.1)]= 2\times10^{-6}$
and $3\times10^{-3}$ for the bottom two entries in Table 1
respectively. For interstellar dust emission,
$max[L(\Delta_{rms},1.6)] = 4\times10^{-3}$ and $2\times10^{-5}$
for the same two
entries. Clearly these possibilities are strongly ruled out, in
the worst case at $3\sigma$ ($L= 0.003$ for a normal distribution).
In W1, $\Delta_{rms}$ from  extra-galactic sources is expected to
be $\approx 9~\mu$K.  A conservative index for these sources is
$\beta\approx -2$; thus, this explanation too is disfavored. An
anisotropy in the CMB, $\beta = 0$, is the most compelling
interpretation of the data.

	In the analysis, we assume that only one celestial
component produces the anisotropy. Dodelson \& Kosowsky (1994b)
have performed a  more general analysis on the SK93 data that
allows two fluctuating components, CMB and free-free. The result
was a decrease in the significance  of the detection of CMB. With
this new limit on $\beta$,  $\Delta_{rms}$ of uncorrelated
free-free emission \footnote{The free-free (or dust) contribution
is estimated by assuming that it adds in quadrature to the CMB
signal. We assume
$\sigma_{CMB}/\sigma_{ff}=$$(\beta_{ff}-\beta)/(\beta_{CMB}-\beta)$.
Then, $\Delta^{2}_{rms} = \sigma_{CMB}^2 + \sigma_{ff}^2 $. For
$\Delta_{rms} = 44~\mu$K and $\beta=-0.9$, $\sigma_{ff}=27~\mu$K,
and $\sigma_{CMB}=35~\mu$K, near the bottom of our confidence
range.} is less than $\approx 27~\mu$K, an additional constraint
on two component models. Using these data alone, we cannot rule
out a model which assumes co-located interstellar dust and free-free
emission, but, based on FIRS (Ganga 1994) we expect dust to
contribute less than $1~\mu$K, again, added in quadrature.

	Gaussian correlation functions are often used to compare
measurements. We find $C_0 = 60^{+14}_{-11}$ for
a correlation angle $\theta_{c}=1.0^{\circ}$.
Alternatively, the result may be interpreted in the context of
the `standard' CDM model (Steinhardt 1993; Bond et al. 1994),
($\Omega_b = 0.05, H_{0} = 50~{\rm km/s/Mpc}, \Lambda =
0$, $n_s=1$, $\Omega = 1$, no reionization), for which we find
that the root-mean-squared quadrupole has a value
$Q = 18.9^{+5.4}_{-3.6}~\mu$K.

%

\acknowledgments

	We would like to thank George Sofko and Mike McKibben at
the The University of Saskatchewan and Larry Snodgrass at the
Canadian SRC for their valuable assistance during the  1994
observing season. We are indebted to Marian Pospieszalski and
Mike Ballister of NRAO  who provided the amazing HEMT amplifiers
that made this experiment possible. We also thank  Chris Barnes,
Carrie Brown, Weihsueh Chiu, Randi Cohen, Peter Csatorday, Cathy
Cukras, Peter Kalmus, John Kulvicki, Wendy Lane, Young Lee, Naser
Quershi, and Peter Wolanin for building many of the components of
the experiment. This work was supported by NSF grant PH 89-21378,
NASA grants NAGW-2801 \& NAGW-1482, and and NSF NYI grant to L.
Page.

	The data, analysis software, beam profile, and correlation
coefficients will be made publicly available upon acceptance of
this {\it Letter}.

%
%
%
%

\clearpage

%



\makeatletter
\def\jnl@aj{AJ}
\ifx\revtex@jnl\jnl@aj\let\tablebreak=\nl\fi
\makeatother


\begin{planotable}{lrrrrcrrrrr}
\tablewidth{40pc}
\tablecaption{Summary of Results}
\tablehead{
\colhead{ Data Set}                         &
\colhead{ Total time\tablenotemark{a}}      &
\colhead{ $\zeta$\tablenotemark{b}}        &
\colhead{ Mean offset\tablenotemark{c} }    &
\colhead{ $\Delta_{rms}$\tablenotemark{d} } &
\colhead{ $\beta$ }                         &
\colhead{ $\delta T_l$\tablenotemark{e} }   \\
\colhead{ }                                 &
\colhead{ (h)   }                           &
\colhead{ (mK/deg)}               &
\colhead{ ($\mu$K) }                        &
\colhead{ ($\mu$K)}                         &
\colhead{ }                                 &
\colhead{ ($\mu$K) }
}
\startdata

${\rm K_a93}$\tablenotemark{f}\dotfill      &
$130 $                              &
$ 4.5$                              &
$-60/-360$                          &
$37^{+12}_{-9}$                     &
$-0.28^{+0.79}_{-1.06}$             &
$36^{+12}_{-9}$                     \nl
                                    &
$ 79$                               &
$ 3.0$                              &
\nodata                            &
$44^{+14}_{-10}$                    &
$-0.69^{+0.74}_{-1.16}$             &
$42^{+13}_{-10}$                    \nl
${\rm K_a93},~$A-B\tablenotemark{g}\dotfill        &
$130$                               &
$4.5$                               &
\nodata                            &
$0,31$                           &
\nodata                            &
\nodata                            \nl
${\rm K_a94}$\dotfill               &
$140$                               &
$2.5 $                              &
$+110/-340$                         &
$49^{+12}_{-7}$                     &
$-0.70^{+0.63}_{-0.74}$             &
$47^{+12}_{-7}$                     \nl
                                    &
$107 $                              &
$1.7 $                              &
\nodata                            &
$52^{+14}_{-8}$                     &
$-0.77^{+0.65}_{-0.83}$             &
$52^{+13}_{-8}$                     \nl
${\rm K_a94},~$A-B$^{\rm g}$\dotfill &
$140 $                              &
$2.5 $                              &
\nodata                            &
$0,41$                              &
\nodata	                    &
\nodata                            \nl
${\rm Q94}$\dotfill                 &
$158 $                              &
$2.5$                               &
\nodata$/-400$                      &
$61^{+17}_{-12}$                    &
$+1.35^{+1.4}_{-2.74}$             &
$57^{+16}_{-11}$                  \nl
                                    &
$98 $                               &
$1.7$                               &
\nodata                            &
$52^{+19}_{-18}$                    &
$-5.97^{+7.23}_{<-5.}$                    &
$48^{+18}_{-17}$                  \nl
${\rm K_a93 - K_a94}$\dotfill       &
$ 298  $                            &
$4.5, 2.5$                          &
\nodata                            &
$0,33$                          &
\nodata                            &
\nodata                            \nl
${\rm K_a93 + K_a94}$\dotfill       &
$ 298$                              &
$4.5$, $2.5$                          &
\nodata                            &
$44^{+11}_{-6}$                     &
$-0.43^{+0.48}_{-0.54}$             &
$42^{+11}_{-6}$                     \nl
                                    &
$186$                               &
$3.0$, $1.7$                          &
\nodata                            &
$49^{+12}_{-7}$                     &
$-0.62^{+0.48}_{-0.59}$             &
$47^{+12}_{-7}$                     \nl
${\rm K_a93 + K_a94},~$A-B$^{\rm g}$\dotfill&
$298$                               &
$4.5$, $2.5$                          &
\nodata                            &
$0,34$                         &
\nodata                            &
\nodata                            \nl
${\rm K_a94 + Q94}$\dotfill         &
$298$                               &
$2.5$, $2.5$                          &
\nodata                            &
$48^{+11}_{-7}$                     &
$+0.10^{+0.44}_{-0.56}$             &
$44^{+10}_{-6}$                     \nl
                                    &
$205$                               &
$1.7$, $1.7$                          &
\nodata                            &
$51^{+13}_{-8}$                     &
$-0.54^{+0.56}_{-0.69}$             &
$47^{+12}_{-7}$                     \nl
${\rm K_a93 + K_a94 + Q94}$ &
$428$                               &
$4.5$, $2.5$, $2.5$                     &
\nodata                            &
$44^{+11}_{-6}$                     &
$+0.17^{+0.42}_{-0.45}$              &
$42^{+11}_{-6}$                     \nl
                                    &
$ 284 $                             &
$3.0,~1.7,~1.7$                    &
\nodata                            &
$47^{+12}_{-7}$                     &
$-0.40^{+0.44}_{-0.52}$             &
$44^{+11}_{-6}$                     \nl

\end{planotable}
\clearpage

\begin{enumerate}

\item[$^{a}$] Between February 21 and March 15 of 1993, the
sky was observed with the ${\rm K_a93}$ radiometer for 450 hours.
Between January 13, 1994 and February 28, 1994, the sky was
observed for 243 hours with the ${\rm K_a94}$ radiometer data,
and 363 hours with the ${\rm Q94}$ radiometer.

\item[$^{b}$] The cut level, $\zeta$ is the root-mean-squared
amplitude  in the horizontal temperature  gradient
(single-difference) for 1s of data. The larger cut levels
correspond to factors of  3.1, 2.5, and 2.5 above system noise
($T_{sys} = T_{rec} + T_{atm}$)  noise for the ${\rm K_a93 }$,
${\rm K_a94}$, and ${\rm Q94 }$ radiometers respectively. The
single difference in the 1994 observing scheme is much more
sensitive to atmospheric gradients than that from 1993.  Data
from the single-difference are not used in the anisotropy
analysis. The data from the double-difference, which we analyze
for anisotropy,  are dominated by the system noise.

\item[$^{c}$] The offsets are given as (vertical
polarization, A)/(horizontal polarization, B).

\item[$^{d}$] Data are quoted as the maximum of the likelihood with the range
of the
error bars encompassing the 16\% to 84\%  integral of the
likelihood marginalized with a uniform prior. Where there is no
``$2\sigma$'' lower bound, the maximum and  95\% upper bound are
given. The units are thermodynamic temperatures relative to a
2.726 K Planck emitter. At 30 GHz, $\partial T_{ant}/\partial T$
$= 0.98$; at 45 GHz,  $\partial T_{ant}/\partial T$ $= 0.95$.
Given our finite sampling of the CMB, presumed to be a random
field, it is not possible to determine $\Delta_{rms}$ to better
than $\approx\pm~6.5~\mu$K.

\item[$^{e}$] The synthesized  beams differ
by about 4\% between channels. For ${\rm K_a93}$ the window spans
$l_{eff}=66^{+27}_{-20}$ and $\sqrt{I(W)}=1.04$ (Bond, 1994); for ${\rm K_a94}$
the window spans $l_{eff}=67^{+29}_{-20}$ and $\sqrt{I(W)}=1.04$; and
for ${\rm Q94}$ the window spans $l_{eff}=71^{+28}_{-23}$ and
$\sqrt{I(W)}=1.08$. $\Delta_{rms}=\delta T_l/\sqrt{I(W)}$. For
the $K_a$ and Q combination, we use $l_{eff}=69^{+29}_{-22}$ and
$\sqrt{I(W)}=1.06$.

\item[$^{f}$] The values for the
$\rm{K_a93}$ data have been recomputed given our increased
understanding of Cas-A and its environs. The original values were
$\Delta_{rms}=33\pm10~\mu$K and $\beta=-0.3^{+0.7}_{-1.2}$. The
cut level $\zeta=4.5~$mK/deg corresponds to
$\eta=3$ in W1.

\item[$^{g}$] This is a measure of the
anisotropy of the difference between two perpendicular
polarizations. It is also a gauge of the consistency between the
A and B channels. The values are for $\beta=0$.

\end{enumerate}

\clearpage

\clearpage

\begin{figure}
\centerline{\epsfxsize=5in\epsfbox{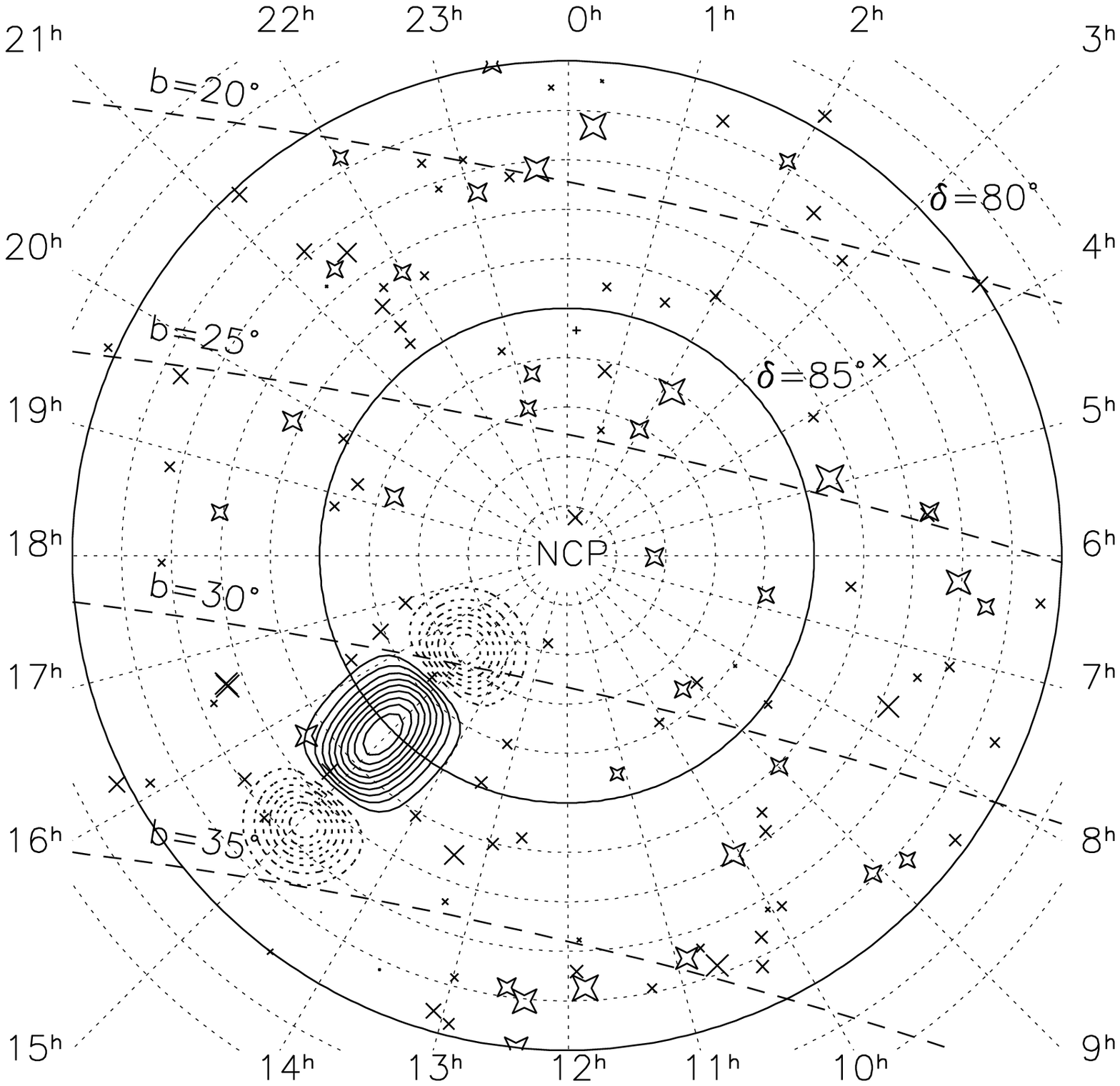}}
\caption{The region of observation and beam pattern for
SK93.  The contours show the beam profile for 1s of data. In the
analysis, all the data with the beam center within one hour of
right ascension are combined; thus, the above beam pattern is
smeared out. Note how much the pattern resembles a
``double-difference'' (positive lobe:solid lines; negative lobes:
dashed lines). The `+' signs mark the  sources in K\"uhr et al.
(1981). Stars mark the  sources with flat spectra. The
size of the symbol is proportional  to the flux. Lines of
constant galactic latitude are also shown. Note the slight
curvature of the observing pattern with respect to  lines of
constant longitude; this is a result of the azimuthal chop. Note
that if there were a hot spot in the CMB near  $\delta =
87^{\circ}$ and
$\alpha = 5^{h}$, the data in bins 4 through 8 would tend to be low.}
\end{figure}

\begin{figure}
\centerline{\epsfxsize=5in\epsfbox{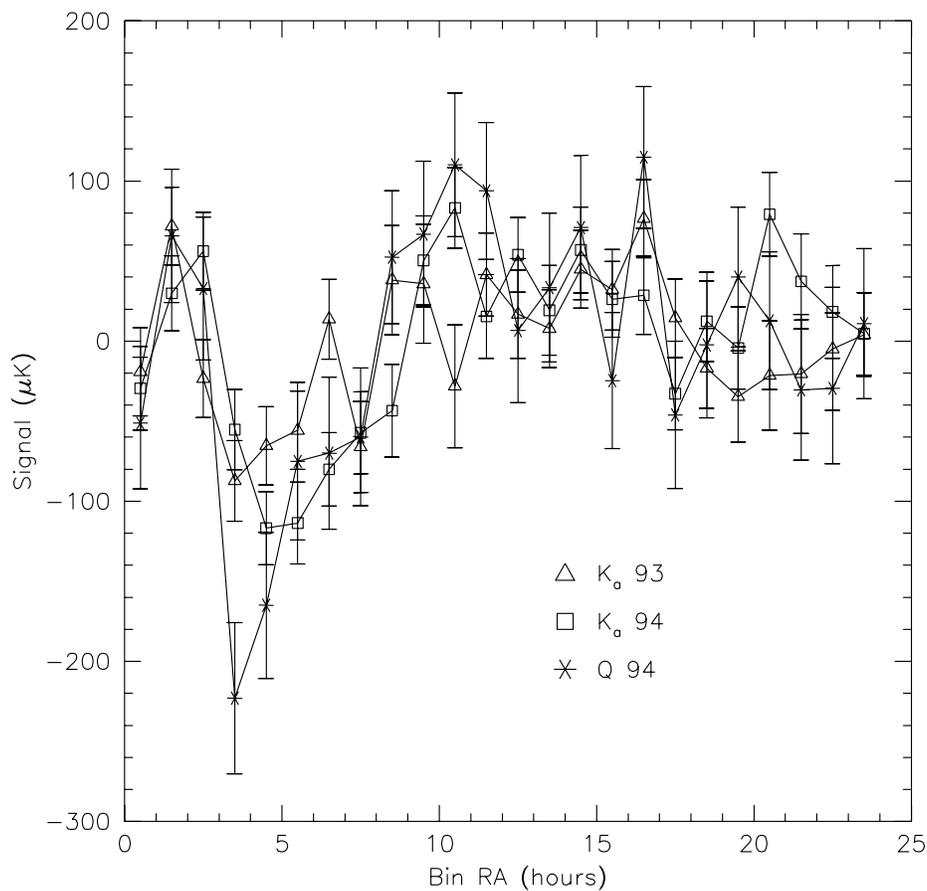}}
\caption{The data for ${\rm K_a93}$, ${\rm K_a94}$, and
${\rm Q94}$ as a function of right ascension bin. Data from the
six ${\rm K_a}$ bands have been added together as have the data from the
three Q bands. Also, the data taken in the east have been
combined with the  data taken in the west. The error bars include
the effects of the correlations between channels. There are no
correlations, other than those from the CMB, between the three
data sets. Because of the observing strategy, each bin is
correlated with the adjacent bins. All correlations have been
accounted for in the analysis.}
\end{figure}

\end{document}